\documentclass[vci-paper]{vciart}
\usepackage{amssymb}
\usepackage{epsfig}

\begin{document}

\vspace*{-1.8cm}
\begin{flushright}
{\large\bf LAL 04-15}\\
{\large\bf DAPNIA 04-79}\\
{\large\bf LBNL-54789}\\
%\vspace*{-0.5cm}
\vspace*{0.1cm}
{\large May 2004}
\end{flushright}

\begin{frontmatter}

\centering
\title{First Tests of a Micromegas TPC in a Magnetic Field}

% if there is only one institution, use this form:
%\author{John Author, Giovanna Autore}
%\address{University of Wisdom, Physics City, Scienceland}

% else, use optional labels to link authors explicitly to addresses,
% as shown below:
\author[1]{Paul Colas}, 
\author[1]{Ioannis Giomataris}, 
\author[2]{Vincent Lepeltier},
\author[3]{Michael Ronan}
\address[1]{DAPNIA, CEA Saclay, 91191 Gif sur Yvette cedex, France}
\address[2]{LAL Orsay, IN2P3-CNRS et Universit\'e de Paris-Sud,
91898 Orsay cedex, France}
\address[3]{LBNL Berkeley, CA, USA}

\begin{abstract}
Since the summer of 2003, a large Micromegas TPC prototype 
(1000 channels, 50 cm drift, 50 cm diameter) has been operated in a 
2T superconducting magnet at Saclay. 
A description of this apparatus and first
results from cosmic ray tests are presented.
Additionnal measurements using simpler detectors with
a laser source, an X-ray gun and radio-active sources are discussed.
Drift velocity and gain measurements, 
electron attachment and aging studies for a Micromegas TPC are presented. 
In particular, using simulations and 
measurements, it is shown that an Argon-CF${}_4$ mixture 
is optimal for operation at a future Linear Collider.
\end{abstract}
\end{frontmatter}

\section{Introduction}
The European TESLA detector and American Large
Detector designs include a large Time Projection Chamber
(TPC) for the main tracking device.
The efficient reconstruction of collimated jets expected at high 
energy $\rm{e^+e^-}$ colliders
requires excellent two-track separation and full 
coverage of endplane detectors using few-millimeter-width anode pads.
To reduce the effect of the severe background conditions, 
fine granularity readout and low gas sensitivity
to neutrons (e.g. a Hydrogen-less mixture) are required, and the
chamber must work in a magnetic field of 3 to 4~T.
We proposed Micromegas\cite{yannis} for the amplification stage of the TPC 
to meet these requirements and to provide
a natural suppression of the ion backflow into the drift 
volume~\cite{vincent}.

The Berkeley-Orsay-Saclay cosmic-ray setup and the data taking
are described in Section 2. In Section 3, we show the Monte-Carlo
simulations and experimental studies which 
favour a new mixture for use in such a TPC. 
In Section 4 new developments on the expected spatial resolution are 
addressed.

\section{The Berkeley-Orsay-Saclay cosmic ray setup}
A 50 cm drift length and 50 cm diameter TPC, equipped with 1024
electronic channels, has been in operation since July 2003, and has taken
magnetic field data in November 2003. The anode is segmented in 8 lines with 96 pads ($2 \times 10 \, \rm{mm^2}$), plus 2 central 
lines of 128 narrower pads ($1 \times 10 \, \rm{mm^2}$). The pads are
drawn on a Printed Circuit Board (PCB) by standard etching technics.
Each pad is read out independently through a metallised via across
the PCB. Particular care has been given to the surface quality of the 
PCB. To sustend the micromesh, 50 $\mu$m high polyimide pillars, 
200 micron diameter, have been formed by etching a photoimageable
film. The large copper micromesh has been made at the CERN workshop.
The detector is fitted to the bore of a 2~T superconducting magnet
providing a magnetic field, homogeneous at the percent level in the
region $\pm 25$ cm each side of the center of the magnet, along the
magnet axis. 

The signals are amplified, shaped and sampled at the rate of 
20 MHz and digitised over 10 bits with the STAR readout 
electronics.
This is the largest micropattern detector ever built.
The VME-based data acquisition is triggered at a rate of about 1 Hz 
by the coincidence of 2 
large scintillators read out by phototubes. The data acquisition 
conditions were very steady, with mesh currents less than 0.3 nA, 
no sparking. The data were analysed 
using Java Analysis Studio and AIDA. Data have been taken with three gas 
mixtures: $\rm{Ar/CH_4 : 90/10}$,
$\rm{Ar/Isobutane : 95/5}$,
$\rm{Ar/CF_4 : 97/3}$.
The drift field was 120 V/cm in the first
case and 200 V/cm in the two others. 
Most tracks were cosmic muons with momentum between 300 MeV and 3 GeV, thus
minimum ionising and with relatively low multiple scattering.

\begin{figure}                                                    
\center                                                          
\epsfig{figure=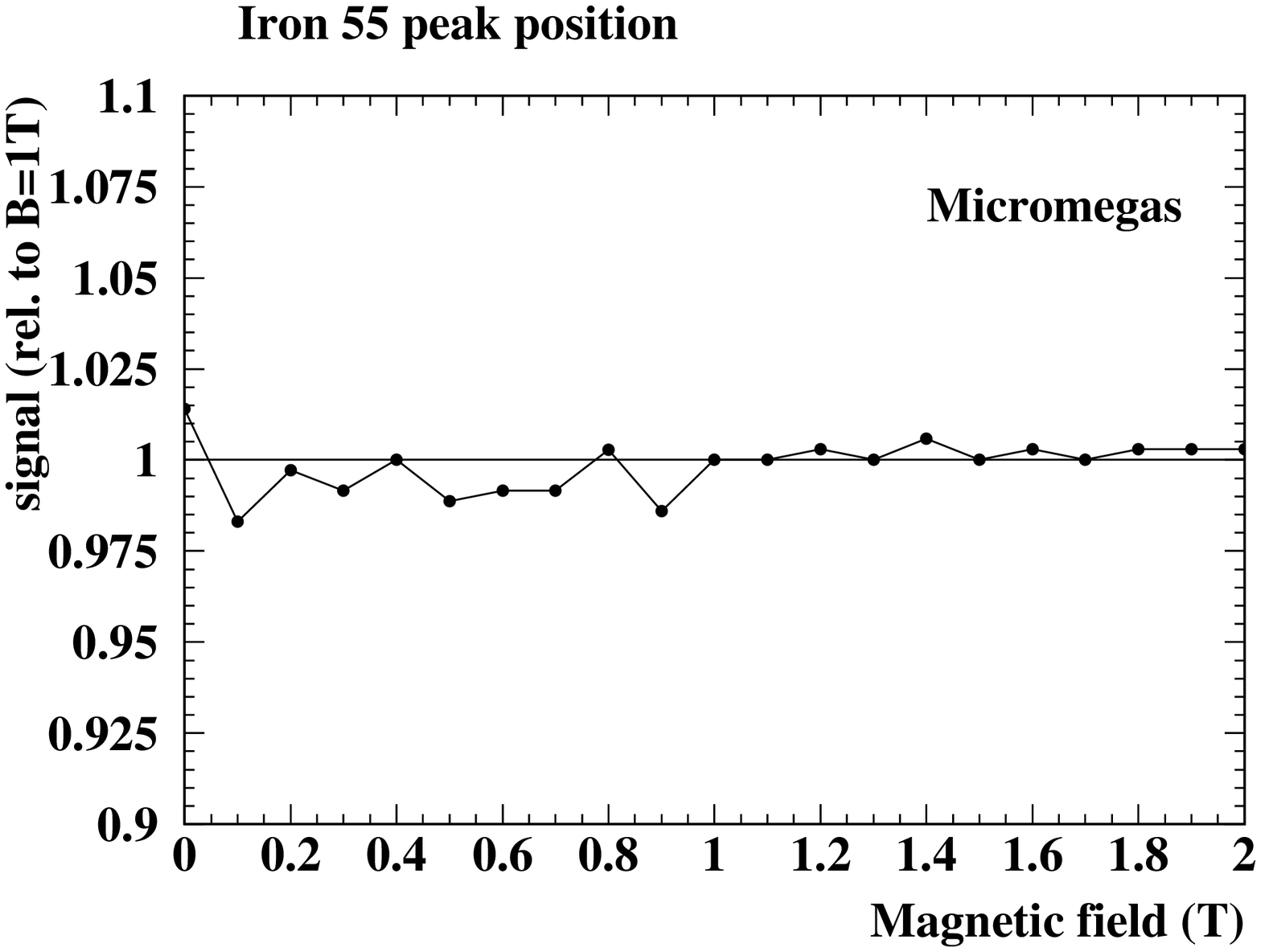,height=10cm}
\epsfig{figure=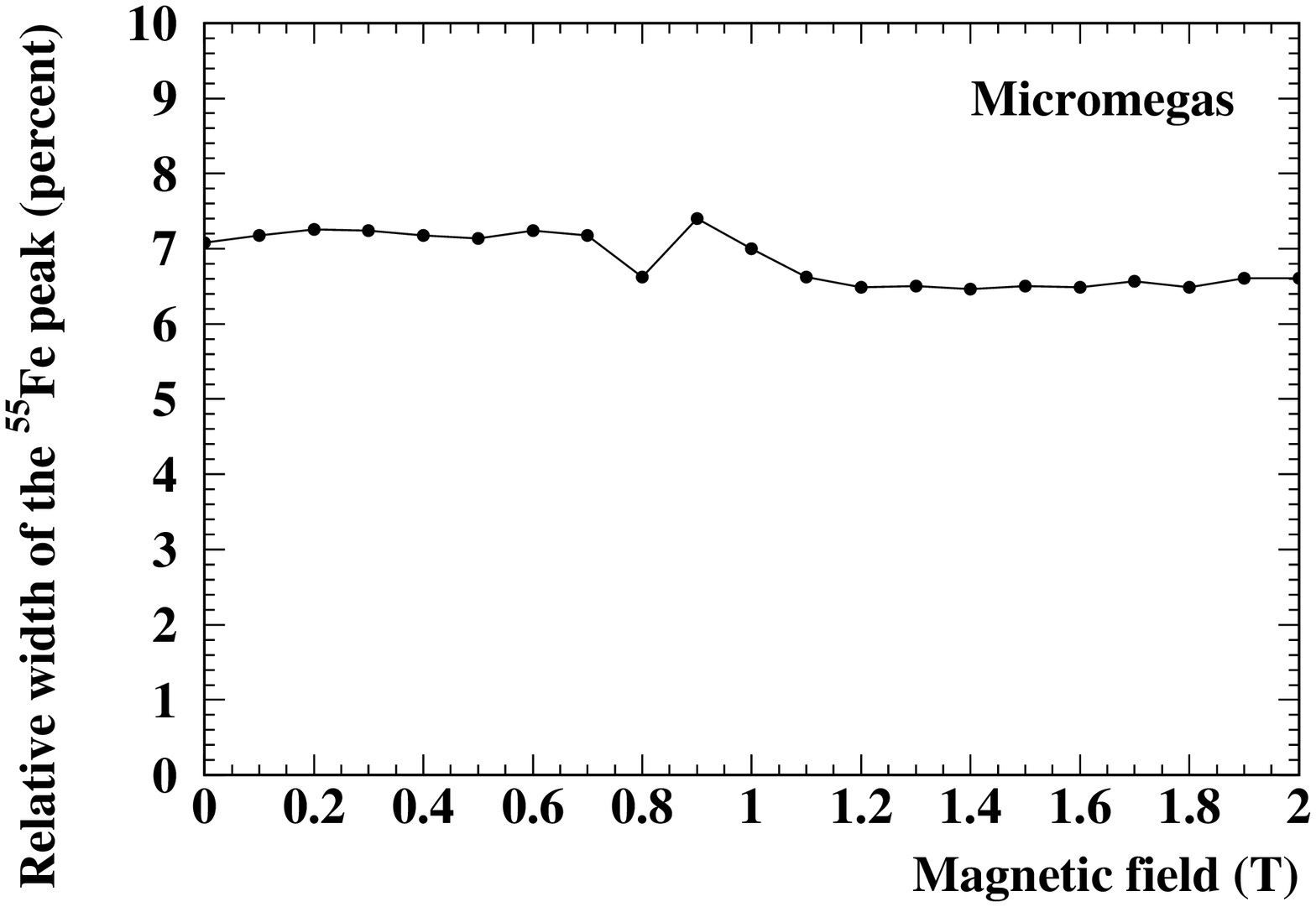,height=10cm}             
\caption{The peak position relative to B=1T (top) and the width (bottom) 
of the $\rm{^{55}Fe}$ 5.9 KeV X-ray
line measured with a Micromegas, as a function of the magnetic field.
}
\label{fig:Fig1}
\end{figure}

Prior to building the large setup, 
we have checked with a small detector with a non-segmented anode
that the behaviour of Micromegas was not hampered
by a large magnetic field perpendicular to the mesh. We measured
the position and the relative width of the 5.9 keV line of an 
iron 55 source, while varying the magnetic field from 0 to 2\,T.
They show a remarkable stability as a function of the magnetic field~(Figure \ref{fig:Fig1}).

\vspace*{0.5cm}

\boldmath
\section{A magic gas mixture: $\rm{Ar + CF_4}$}
\unboldmath

\begin{figure}[h]                                                   
\center                                                          
\epsfig{figure=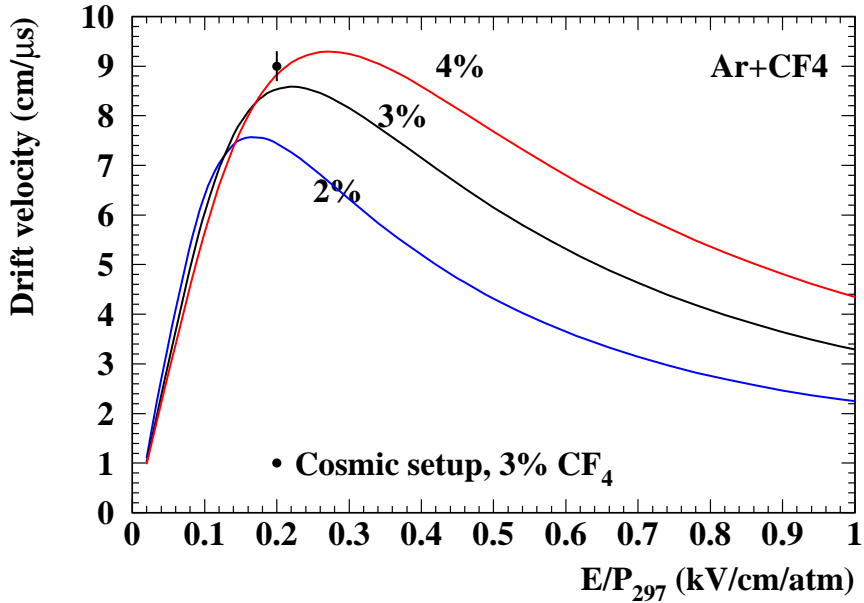,height=9cm}                       
\caption{The drift velocity of electrons as a function of the drift 
electric field for Ar mixtures with 2, 3 and 4\% of $\rm{CF_4}$, 
as predicted by Magboltz. The data point is the measurement with the cosmic 
ray setup, with 3\% of $\rm{CF_4}$.}   
\label{fig:Fig2}
\end{figure}

Beside demonstrating the feasability and operability of a 
large scale Micromegas TPC, the goal of the cosmic ray data taking 
was to confirm Magboltz~\cite{biagi} Monte Carlo calculations 
of the expected performance of different gas mixtures. 
Over 50 workable mixtures of a majoritary noble gas (the ``carrier")
with an admixture (the ``quencher") of one or two
molecular gases to quench the UV photons produced during the avalanche,
have been considered.
Simultaneously requiring to have enough primary electrons, have an affordable
cost, and present a velocity maximum at low enough field -- a field 
of 200 V/cm already requires a 50 kV cathode voltage for a drift length of 
2.5 m -- point to Argon as a carrier gas. Most of the quenchers used until now
are hydrocarbons. However, Hydrogen nuclei, protons, are bounced by 
O(1 MeV) neutrons that are expected to be produced in large numbers by the 
accelerator and heavily ionise while drifting around magnetic field lines.
It is thus preferable to avoid hydrogenated gases. $\rm{CO_2}$ gives 
too small drift velocities in mixtures with Ar. In contrast, a few percent of 
$\rm{CF_4}$ gives velocities up to 8-9 $\rm{cm/ \mu s}$ at a low enough field
as shown by the curves of Figure  \ref{fig:Fig2}.

{\bf Drift velocity.}
With the mixture of Ar with 3\% $\rm{CF_4}$, the time between the trigger 
and the arrival of the ionisation buckets on each time cluster was determined.
The time distribution presents an edge at
5350 ns with a 3\% accuracy, for a 47.9 cm drift. This leads to a drift 
velocity determination of $9.0 \pm 0.3 \, \rm{cm/ \mu s}$ in agreement with 
the Magboltz prediction of $8.6 \, \rm{cm/ \mu s}$.
As a cross-check, a drift velocity of $4.24 \pm 0.10 \, \rm{cm/ \mu s}$ 
was measured for an Ar + 5\% isobutane mixture, where the expectation is 
$4.16 \, \rm{cm/ \mu s}$.

\begin{figure}[h]                                                    
\center                                                          
\epsfig{figure=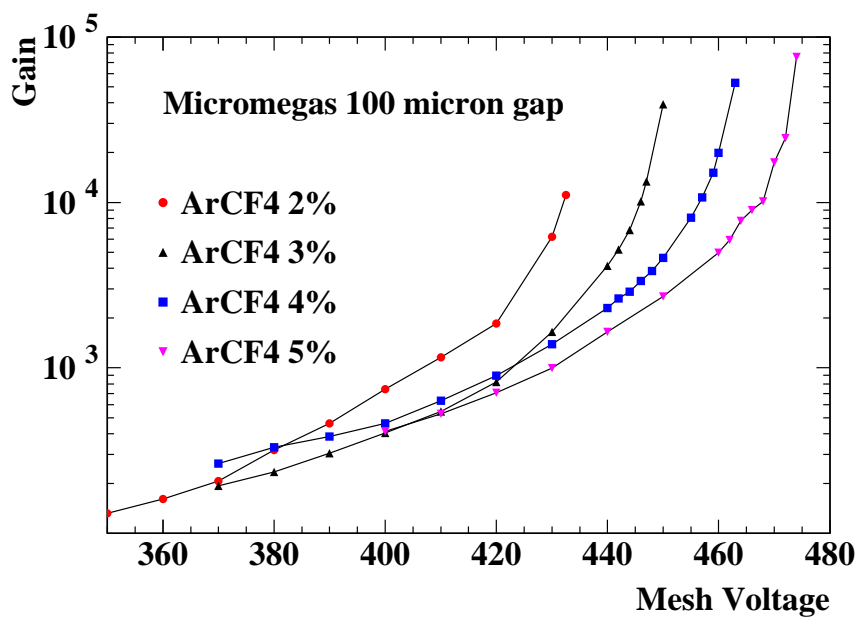,height=9cm}                       
\caption{Gain as a function of the mesh voltage, measured with a 
$\rm{^{55}Fe}$
source and a calibrated amplification chain, for various percentages of 
$\rm{CF_4}$ in Argon.}  
\label{fig:Fig3}
\end{figure}

{\bf Gain.}
The gain has been measured as a function of the mesh voltage for various 
concentrations of $\rm{CF_4}$ in Ar, with a 100 $\rm{\mu m}$ gap. 
The result is shown in figure \ref{fig:Fig3}. Gains up to $10^5$ are reached before 
sparking. The gain shows an exponential behaviour until a value of 1000,
due to secondary effects, 
and grows faster than an exponential for higher voltages. 
With existing or foreseeable low-noise electronics, 
gains as low as 300 or 500 would suffice to operate the detector, and would 
even be seeked to avoid the formation of space charge in the drift volume.
In the cosmic setup, data were taken at a gain of about 800. 

{\bf Attachment.}
There exists a resonance in the attachment cross-section for an 
electron kinetic energy slightly below the ionisation 
threshold~\cite{christophorou}.
This might hamper the operation of a device using this gas, as the
negative ions formed are excessively slow. However, the Monte Carlo
simulation 
predicts no attachment (or a negligible attachment) in the drift region,
for fields less than 400 V/cm, and also predicts that the attachment
is overwhelmed by the Townsend coefficient in the amplification region.
Given the smallness (a few microns) of the transition between the two
regions, the operation of Micromegas is not affected by attachment.
In the cosmic ray setup, 
an exponential fit to the truncated mean signal versus drift distance  
showed no attenuation, allowing a lower limit at 2.4 m (90\% C.L.) to be set
on the attenuation length.

\begin{figure}[h]                                                    
\center                                                          
\epsfig{figure=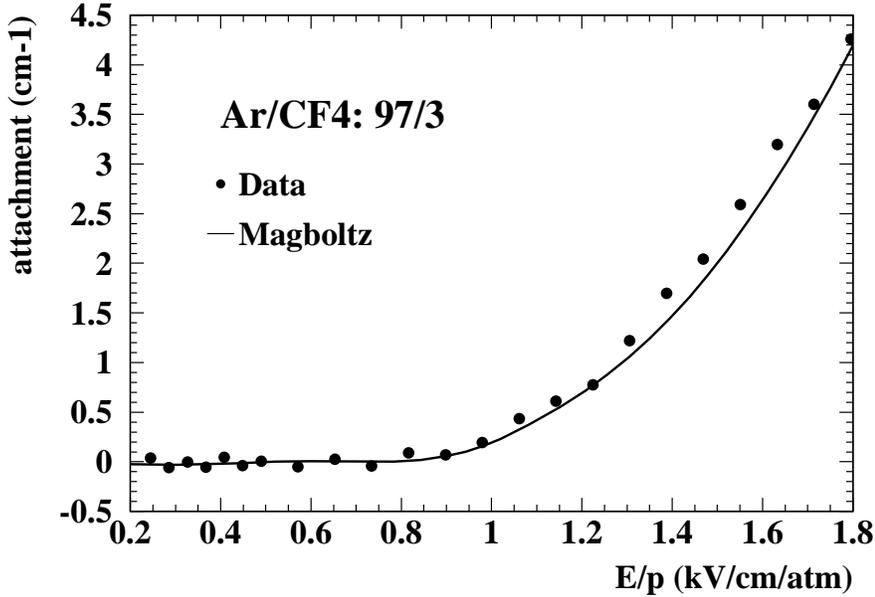,height=9cm}                       
\caption{Attachment coefficient of an Argon (97\%) $\rm{CF_4}$ mixture 
as a function of the electric field. The points are data described 
in the text and the line is the Magboltz Monte Carlo prediction.}
\label{fig:Fig4}
\end{figure}

The dependence of the attachment coefficient as a function of the electric
field has been measured from the 
amplitude of a laser photoelectric signal, in a setup with a 1.29 cm drift.
The data are in excellent agreement with Magboltz predictions (Figure \ref{fig:Fig4}).

{\bf Aging.}
Mixtures containing $\rm{CF_4}$ have often been convicted of damaging the
detectors flushed with them, 
especially in the presence of Hydrogen atoms and in highly ionising 
conditions~\cite{aging}.  
In the past 5 years however many tests have been carried out with 
Micromegas detectors with $\rm{CF_4}$ mixtures, and never signs of degradation
have been observed. A dedicated aging test has been carried out at Saclay with
an Ar plus 5\% $\rm{CF_4}$ mixture, by monitoring the mesh current during 
3 days, gathering $\rm{4 \, mC/mm^2}$ (over 1,000 years
of linear collider with expected background conditions!). The gain was 8000 
and no degradation was observed at the percent level. 
Though the aging issue would require additionnal long-term systematical 
studies, there is no reason to
fear anomalous aging, in well-controlled operating conditions and with 
a suitable choice of the building materials. 

\begin{figure}[h]                                                   
\center                                                          
\epsfig{figure=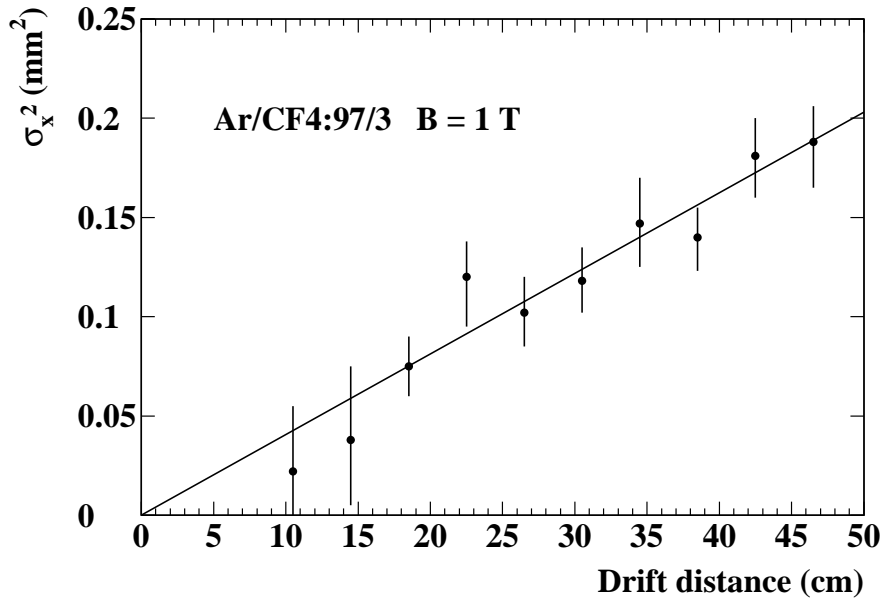,height=9cm}                       
\caption{Track r.m.s. width measured at 1T as a function of the
drift distance.}
\label{fig:Fig5}
\end{figure}

{\bf Transverse diffusion.}
A circle has been fitted to the projection of
each track onto the plane transverse to the electric and magnetic fields.
Six out of the ten pad rows (including the two with narrow pads) are used
in the fit, while the r.m.s. width of the 4 leftover hits is estimated 
from a maximum likelihood fit to the amplitude distributions as a function
of the coordinate along the pad row. 
Most information comes from the hits consisting of more than 2 pads
(50\% of the pads are such) and from tracks near the edge of a single-pad
hit.
 The square of the average of the r.m.s. hit widths ($\sigma_x^2$) for 
the 1 tesla data sample are plotted versus the drift distance in figure \ref{fig:Fig5}.
The linear increase of $\sigma_x^2$ as a function of the drift distance
is characteristic of the diffusion. The slope gives the transverse diffusion 
constant for Ar plus 3\% $\rm{CF_4}$ at B=1\,T: 
$\rm{D_t=64 \pm 16 \, \mu m / \sqrt{cm}}$ where the error is dominated
by systematics, at this preliminary stage of the analysis. 
This is in agreement with the
expectation of $\rm{86 \, \mu m}$. Note that the $\omega \tau$ factor 
for this fast gas for B = 1~T is as large as 4.  

\section{Spatial resolution}
The very low transverse diffusion expected for a LC-TPC operating in 
magnetic fields of 3-4~T yields a potential for a very high 
spatial resolution. 
However, there is a drawback: at a typical drift distance 
of 1\,m, the r.m.s. width of a track is expected to be $\sim 350 \, \mu$m, 
much smaller than the 2~mm-wide pads that would provide the required
2-track separation. 
The charge would be collected by a single pad, yielding resolutions of 
order $600 \, \mu$m, much worse
than one could expect if one could make a barycenter between 2 or 3 
neighbouring pads. 
Narrower pads would imply an unacceptable increase in the the number 
of electronic channels. Two solutions have been proposed to this problem. 

One possibility is to spread the charge after amplification, by means for 
instance 
of a resistive foil~\cite{madhu}. This has been proven to function with 
GEMs and 
Micromegas and to yield point resolutions of $\rm{80\,\mu m}$, 
beyond the requirements for the LC TPC.

Another possibility is the digital TPC~\cite{hauschild}:
$\rm{300 \times 300 \, \mu m ^2}$ pads equipped with a digital readout 
would provide the ultimate resolution and a better dE/dx measurement.
Practical demonstrations of such devices with a gas amplifier 
combined with silicon pixels have been  
presented for the first time at this conference~\cite{ronaldo,harry}. 
This would require $10^8$ channels, but all integrated and with a binary
(i.e. 1 bit) output. 

\section{Conclusion}
A large Micromegas TPC has been operated successfully for the first time in a
magnetic field. The first results show that an Ar $\rm{CF_4}$ mixture
is particularly suited for operation at the linear collider. 
Obtaining the optimal spatial resolution is still a challenge, 
but two satisfactory solutions have been
recently demonstrated in principle. 
At the present stage of the $\rm{R \& D}$,
the adequacy of a Micromegas TPC for the tracking at the linear
collider appears thus very promising.

\section*{Acknowledgements}
We wish to thank F. Bieser, R. Cizeron, C. Coquelet, E. Delagnes,
A. Giganon, G. Guilhem, R. de Oliveira, V. Puill, Ph. Rebourgeard 
and J.-P. Robert 
for their help in building and commissionning the detector, and 
D. Karlen for providing us with the Victoria display and analysis 
software.

\end{document}